# Tight focusing proton beam with radius in nanometer scale generation based on channeled solid target

**Q. Yu[1]***

[1] *School of Mechanical Engineering and Rail Transit, Changzhou University, Changzhou 213164, China*

Abstract

An efficient scheme of generating ultra-tightly focused proton bunch with radius in nanometer scale is proposed. A needlelike proton filament of transverse size in nanometer scale with the density of $23n_c$ and $\sim 3.7nC/\mu m$ charge quantity is obtained based on multi-dimension Particle-in-Cell (PIC) simulations. The regime is achieved via laser irradiating on a solid target with pre-channeled density profile. The theoretical analysis mentions that the transverse electric field dramatically transits from a defocusing dipole to double dipoles structure with the change of the initial target density distribution from uniform to pre-channeled. The inner dipole of the electric field tightly focuses the proton beam into the order of magnitude of nanometer. 3D simulations verify the scheme in the realistic condition. Various pre-channeled density profiles including linear, parabolic and arbitrary steeped prove to work well for the regime, which declares the robustness and the performability of the scheme in experiment.

*Corresponding Author: qin_yu_1@163.com

Owing to the high acceleration gradient, short temporal duration, particle source compactness, and/or high beam density, laser driven plasma accelerator has been considered as an high promising alternative for producing the high-quality charged-particle beams [1-3]. Ion acceleration driven by super-intense laser pulses has undergone impressive advances in recent years. Based on the unique characterastics of high energy ion beam including small device size and high brightness, the laser-driven ion sources have prospective applications in fast ignition of fusion core [4], warm dense matter generation [5], tumor therapy [6], radiography [7], brightness enhancement for conventional accelerators [8] and isotope generation [9].

With the laser intensity increasing, much progress in the understanding of fundamental physical phenomena about ion acceleration has been achieved [10-12]. Different ion acceleration schemes have been described *e.g.* target normal sheath acceleration (TNSA) [13], radiation pressure acceleration (RPA) [14], breakout after burner acceleration (BOA) [15], shock acceleration [16] and magnetic vortex acceleration [17]. At present, the most widely understood and the most stable mechanism is Target Normal Sheath Acceleration. This mechanism [18] and a related radiation-pressure hybrid-scheme [19] can drive protons to energies approaching 100 MeV. However, producing an energetic proton beam with high flux, small emittance, big compactness and low energy spread proved to still be the major challenge for practical applications. To improve the beam quality, much effort has been made. A foil-in-cone target is proposed by Zou *et al.* [20] to enhance laser-radiation-pressure-driven proton acceleration. Psikal *et al.* [21] found a cylindrical target can increase proton energy. Bin *et al.* [22] showed that the front-surface curvature can affect the proton acceleration. Patel *et al.* [23] experimentally observed focused protons from a semispherical shell target. Here, we propose a promising solution for proton focusing to enhance beam charge flux and compactness via the employment of laser-irradiated dense plasma channel. Both 2-D and 3-D PIC simulations declare an ultra-tightly focused and dense needle-like proton beam with nanometer radius and $23n_c$ density is achieved. The obtained energetic proton needle is also highly collimated with small transverse emittance. The unique features of compactness, high charge flux and good collimation lead the beam to be potentially applied in various areas.

We first review the field structure in the ion channel that features in the standard approach to DLA (the direct laser acceleration). In an initially uniform target, the laser beam tends to expel electrons radially outwards, creating a positively charged channel on its path. The corresponding transverse quasi-static electric field is directed away from the axis and, in the 2D case, based on Maxwell equation, it is roughly given by

$$E_y \approx \rho_c(\cos\alpha_2 - \cos\alpha_1 \ d) \quad (1)$$

where $\rho_c$ is the plasma density, $\alpha_1 = \arccos\left(d_0 / \sqrt{(1+y)^2 + d_0^2}\right)$, $\alpha_2 = \arccos\left(d_0 / \sqrt{(1-y)^2 + d_0^2}\right)$ and $d_0 \to 0$. The electrons that are present in the channel are pushed forward by the laser pulse, creating a longitudinal current. In the 2D case, the corresponding quasi-static magnetic field is perpendicular to the simulation plane and is given by

$$\frac{|e|\overline{B_Z}}{m_e \omega c} \approx -\frac{|j_x|}{|e| n_c c}\frac{y\omega}{c} \quad (2)$$

where $j_x$ is the electron current density, $\omega$ is the laser frequency, $e$ is electron charge, $m_e$ is electron mass and $c$ is the light speed. The transverse force induced by these electric and magnetic fields on the background ions is always directed outwards the axis:

$$F_y = |e|\left(E_y - \frac{v_x}{c}\overline{B_z}\right) \approx |e|\left(\frac{\rho_0}{d_0}(\cos\alpha_2 - \cos\alpha_1) - \frac{v_x}{c}\overline{B_z}\right) \quad (3)$$

Considering extremely slight ion motion, the force from magnetic field is negligible for the ions. The corresponding radial force exerted on the ions becomes:

$$F_y = |e| E_y \approx \frac{|e|\rho_c}{d_0}(\cos\alpha_2 - \cos\alpha_1) \quad (4)$$

Now, considering the target is pre-channeled with linear density profile, $\rho = k\,|\,y\,| + \rho_0$ or parabolic density distribution, $\rho = Ay^2$ , the corresponding transverse quasi-static electric field is given by

$$E_y = -\begin{pmatrix} k\ln\left|\dfrac{\sec\beta_2 + tg\beta_2}{\sec\beta_1 + tg\beta_1}\right| + k(\sin\beta_2 - \sin\beta_1) + \dfrac{B_1}{d_0}(\cos\beta_2 - \cos\beta_1) \\ +k\ln\left|\dfrac{\sec\beta_1 + tg\beta_1}{\sec\beta_3 + tg\beta_3}\right| - k(\sin\beta_1 - \sin\beta_3) - \dfrac{B_2}{d_0}(\cos\beta_1 - \cos\beta_3) \end{pmatrix} \quad (5)$$

where $k$ is the linear density gradient, $B_1 = ky + \rho_0$ , $B_2 = ky - \rho_0$ and $\beta_1 = \arcsin\left(-y / \sqrt{y^2 + d_0^2}\right)$ , $\beta_2 = \arcsin\left((1-y) / \sqrt{(1-y)^2 + d_0^2}\right)$ , $\beta_3 = \arcsin\left((-1-y) / \sqrt{(-1-y)^2 + d_0^2}\right)$, or

$$E_y = A\begin{pmatrix} (\cos\theta_1 - \cos\theta_2)\left[-\dfrac{y^2}{d_0} - \dfrac{d}{\cos\theta_1\cos\theta_2} + d_0\right] \\ -2y\ln\left|\dfrac{\sec\theta_2 + tg\theta_2}{\sec\theta_1 + tg\theta_1}\right| - 2y\left(\sin\theta_1 - \sin\theta_2\right) \end{pmatrix} \quad (6)$$

here, $\theta_1 = \arcsin\left(-(y+1) / \sqrt{d_0^2 + (y+1)^2}\right)$ and $\theta_2 = \arccos\left(d_0 / \sqrt{(1-y)^2 + d_0^2}\right)$. The transverse force induced by these electric and magnetic fields on the pre-channeled ions is : $F_y = |e|\left(E_y - \dfrac{v_x}{c}\overline{B_z}\right)$. Considering the ions with small velocities and neglect the force from the magnetic field, the corresponding radial force acted on the linearly pre-channeled ions and parabolic pre-channeled ions are:

$$F_y = |e|\,E_y = |e|\begin{pmatrix} -k\ln\left|\dfrac{\sec\beta_2 + tg\beta_2}{\sec\beta_1 + tg\beta_1}\right| + k(\sin\beta_2 - \sin\beta_1) + \dfrac{B_1}{d_0}(\cos\beta_2 - \cos\beta_1) \\ +k\ln\left|\dfrac{\sec\beta_1 + tg\beta_1}{\sec\beta_3 + tg\beta_3}\right| - k(\sin\beta_1 - \sin\beta_3) - \dfrac{B_2}{d_0}(\cos\beta_1 - \cos\beta_3) \end{pmatrix} \quad (7)$$

$$F_y = |e|\,E_y = |e|\,A\begin{pmatrix} (\cos\theta_1 - \cos\theta_2)\left[-\dfrac{y^2}{d_0} - \dfrac{d}{\cos\theta_1\cos\theta_2} + d_0\right] \\ -2y\ln\left|\dfrac{\sec\theta_2 + tg\theta_2}{\sec\theta_1 + tg\theta_1}\right| - 2y\left(\sin\theta_1 - \sin\theta_2\right) \end{pmatrix} \quad (8)$$

We display the theoretical and numerical transverse electric field profiles induced by the flat ions, linearly channeled ions and parabolic channeled ions by the red and blue lines in Fig. 1(a)-(c). The theoretical results are from Eqs. (1), (5) and (6), respectively. The numerical electric fields are obtained from 2D PIC simulations with the same parameters as the theoretical cases. It is clear the theoretical and numerical electric field profiles are roughly consistent with each other. Being better able to visually observe the transverse electric field structure, we also present the simulative transverse electric field distributions in 2D frame in Fig. 1(d)-(f). The electric field structures are diagnosed after the penetration of the laser from the target and the electrons and ions have been completely separated from each other. It ensures that the transverse electric field is entirely induced by the ions. Fig. 1(d) illustrates a well-known result that the uniform ion structure induces a defocusing dipole structure of the transverse electric field, generating an expelling force (as described by Eq. (4)) for the ions that pushes them outwards the axis of the channel. There are double transverse electric field dipoles with opposite polarities as depicted in Fig. 1 (e) and (f) when the ion density profile is channeled. The outer one is directed away from the axis, and the inner one is directed towards the axis providing a focusing force (as described by Eqs. (7) and (8)) for the background ions in the channel.

Utilizing the focusing field of the double electric field dipoles structure, the ions will be efficiently focused and bunched to improve the beam compactness and reduce the emittance. 2D PIC simulation is performed to illustrate how the ions are focused with such a transverse field in the pre-channeled solid target. The simulation is performed on the electromagnetic relativistic code “EPOCH”. Simulation size is $25\lambda\times70\lambda$ with $2500\times7000$ cells in longitudinal and transverse directions and there are 52 super particles per cell. The target locates between $10 < x(\lambda) < 15$ with uniform density in laser propagation direction. The density linearly increases from $1.26n_c$ to $70n_c$ between $0 < |y(\lambda)| < 30$. The laser spot size is $d_0 = 10\lambda$, pulse duration $\tau = 3.3T_0$ where $T_0$ is the laser period, and laser intensity $I = 1.37\times10^{21}W/cm^2$. The pulse wavelength is $\lambda = 1.0\mu m$ and laser incidents the plasma from the left surface of the target.

We know the pre-channeled ion density structure generates double dipoles structure of the electric field based on above analysis. The outer one is defocusing and provides a restoring force for hot electrons to prevent them escaping from the target transversely. The forces applied on the hot electrons are balanced between the dipolar electric field (focusing force to the electrons) and the laser transverse electric field (defocusing force to the electrons) to form a hollow free of electrons when the self-channeling condition is satisfied. After the laser penetrates the target, the protons within the hollow start to focus and bunch at $40T_0$ as seen in Fig. 2(a). An ultra-tightly focused proton filament is produced at $65T_0$ as drawn in Fig. 2(b). The proton filament includes $\sim 3.7nC/\mu m$ charge, duration of about $20\lambda$ and it achieves a current of $6.9\times10^{10}A/\mu m$. The proton filament longitudinally spreads away from the center to both sides, just like a needle. The transverse density cut along $y$ in the target center ( Fig. 2(c)) illustrates that the proton needle contains a high density of $20n_c$ and a tiny diameter of $55nm$ at the beginning and then both enhance to $23n_c$ and $300nm$ at $65T_0$ as demonstrated in Fig. 2(d). The proton filament is long-time living because it remains to keep the solid density and the diameter on the order of magnitude of $nm$ for at least $110T_0$ as displayed in Fig. 2(f).

The cut-off energy of the proton filament is ~40MeV. The angular divergence and the transverse emittance are $6^o$ and $0.007mm\cdot mrad$ as depicted in Fig. 4, which indicate the filament is well collimated. When protons begin to focus, the pre-channeled density distribution changes, which weakens the density gradient . It diminishes the strength of the focusing force according to Eq. (7), which decreases the ion oscillation velocity, ensuring the good collimation of the filament.

As analyzed above, the focusing dipole structure of the transverse electric field causes the ions to focus to be a filament, but how the proton filament is accelerated?

The breakout after burner acceleration (BOA) scheme [15] and magnetic vortex acceleration mechanism [17] contribute to the proton acceleration. For the employed target, the smallest density being $1.26n_c$, the corresponding pump depletion length $L_{dp} = (\omega^2 / \omega_p^2)\lambda_p a_0 / (3\pi)$ is less than $4\lambda$. But the thickness of the target is $5\lambda$, which seems to be opacity to the laser pulse. However, the laser is compressed during its propagation through the plasma and the effective laser intensity increases from original $a_0 = 30$ to $a_0 = 54$, which ensures the pulse and the hot electrons to penetrate the target by the BOA regime. Protons are accelerated to reach the target rear side by accelerated electrons from BOA regime. Then, when the hot electrons enter into the right vacuum region, they expand rapidly. This leads the corresponding magnetic field to expand (see Fig. 3(a) and (b)) and induces an acceleration electric field (see Fig. 3(c) and (d)) for the protons via the magnetic vortex mechanism. The accelerated protons originate not only from the target rear but also the whole target depth, which is very different from the general proton acceleration regime such as TNSA, BOA and RPA where the accelerated protons originate either from the target rear side (the former two regimes) or the target front side (the latter one scheme).

Three effects contribute to maintain the radius of the obtained proton bunch in the nanometer magnitude. Firstly, the preformed plasma channel nonlinearly focuses the relativistic laser pulse as the effect of a plasma lens. According to theoretical estimate [24], the focused laser radius is $r_0^{'} = (a_0 n_c / n_e)^{1/2} \lambda / \pi \approx 1\lambda$. The simulation result reveals that the laser focuses to $0.5\lambda$ from $5\lambda$, which is even smaller than the theoretical estimate due to the nonlinear effect. The decreased laser radius reduces the radius of the plasma hollow. This limits the maximal radius of the betatron motion for the accelerated ions and thus imposes a cap on the emittance of the ions. A rough estimates of this cap can be found by assuming that the particles are uniformly distributed in a 4D ellipse of the transverse phase space with restriction equation $y^2 + x^2 + (p_y^2 + p_x^2) / \gamma_e^2 k_\beta^2 \le r^2$, where $r$ is the ion sphere radius and $k_\beta = k_p / \sqrt{2\gamma_e}$ is the betatron wave vector. This leads the maximum emittance to be $\in_{\perp.\max} = \gamma_e k_\beta r^2 / 6$. The smaller hollow radius corresponds to smaller ion emittance. Secondly, the ion density gradient decreases when ions start to focus. The decreased density gradient reduces the strength of the focusing electric field, *i.e.*, the proton betatron oscillations radius and the proton filament radius. Lastly, the protons are neutralized rapidly by the transverse return hot electrons, which prevents the protons transverse momentum from continually increasing. However, the neutralization also lowers the efficiency of the proton acceleration. So the cut off energy in our case is not very high. The removal, by the external B field, of electrons that allowed the ion beam to accelerate has been achieved in experiment [25] and it does not increase the ion transverse momentum.

To verify the proton filament generation mechanism in a more realistic condition, we perform 3D simulations where the laser and target parameters are the same as that in 2D case unless otherwise stated. A needlelike proton filament from linearly pre-channeled case is obtained in 3D case as plotted in Fig. 5(a) which proves the proposed regime works well in the reality dimension. To demonstrate the robustness of the regime, we perform another 3D simulation where the solid target has initial parabolic channeled density distribution. The results are presented in Fig. 5(b), in which a proton filament is gained again. Even if the laser intensity is reduced to $a_0 = 8$, we still gain the proton filament as well provided that the corresponding target thickness is reduced. It is difficult to prepare such channeled solid target in experiment. For this purpose, we change the target density profile to be transversely stepped with radial increasing density. The 3D simulation results are presented in Fig. 5(c) where a needlelike proton filament is formed, which demonstrates the performability of the mechanism in experiment.

**Summary**

In conclusion, theoretical analysis proves that a channeled ion density distribution induces the double cylindrically defocusing and focusing transverse electric field structures. The defocusing field providing the restoring force for the hot electrons balances the laser transverse pondermotive force to generate a hollow free of electrons.

Utilizing the focusing field inside the hollow, the protons are focused and bunched to a needlelike filament of ~300$nm$ radius. The proton filament is dense and well collimated with the peak density of $23n_c$ and $0.007 mm \cdot mrad$ transverse emittance. Although the proton energy (~40MeV cutoff energy) is not so satisfactory, it is a promising candidate injecting beam for further acceleration due to its compactness and collimation. The proposed mechanism is robust since it is proved to be efficient with different target density profile or laser intensity. Changing target to transversely stepped density structure being more realistic in experiments, the scheme still well works.

**Acknowledgements**

This work is supported by the project ELI: Extreme Light Infrastructure (CZ.02.1.01/0.0/0.0/15_008/0000162) from European Regional Development. Computational resources were partially provided by ECLIPSE cluster of ELI-Beamlines. This work is also supported by Natural Science Foundation of China under Contract No. 11804348.

## Figure Captions

**Figure 1**

The theoretical transverse electric field profiles ( indicated by the red lines) induced by uniform ions (a), linearly channeled ions (b) and parabolic channeled ions (c) based on Eqs. (1), (5) and (6) where $\rho_c = 0.005n_c$, $\rho_0 = 0$, $k = 0.05n_c$ and $A = 0.05n_c$. The blue lines present the corresponding numerical results for the three cases where the parameters are the same as the theoretical cases. The simulative transverse electric field distributions for uniform ions (d), linearly channeled ions (e) and parabolic channeled ions (f) cases in 2D frame.

**Figure 2**

The proton density structures at $40T_0$ (a), $65T_0$ (b), $90T_0$ (e) and $110T_0$ (f). The density cut along with $x$=12.6$\lambda$ for $40T_0$ (c) and $65T_0$ (d).

**Figure 3**

The magnetic field distributions at $24T_0$ (a) and $30T_0$ (b). The electric field profiles at $24T_0$ (c) and $30T_0$ (d).

**Figure 4**

(a) The angular-energy distribution of protons at $65T_0$. The radial distance is the energy in MeV. The inset displays the protons energy spectrum at $65T_0$. (b) The protons angular spectrum at $65T_0$.

**Figure 5**

The proton density profiles, projected proton density distributions (on *x*-*y* plane and *y*-*z* plane) for linearly channeled case (a), parabolic channeled case (b) and stepped channeled case (c) from the 3D simulations. The insets exhibit the protons energy spectra for the corresponding cases. The simulation size decreases to $15(\lambda)\times 40(\lambda)\times 40(\lambda)$ with the targets placed at $1 < x(\lambda) < 6$ for all cases. The density linearly increases from $1.26n_c$ to $33n_c$ for linearly pre-channeled case (a) and it is $1.26 < n_i(n_c) < 35$ for parabolic channeled case (b). In the stepped channeled case (c), the density changes from $1n_c$ to $68n_c$ with step width of $1\lambda$ and height of $4.5n_c$.

**Figure 1**

**Tight focusing proton beam with radius in nanometer scale generation based on channeled solid target**

**Q. Yu *et al.***

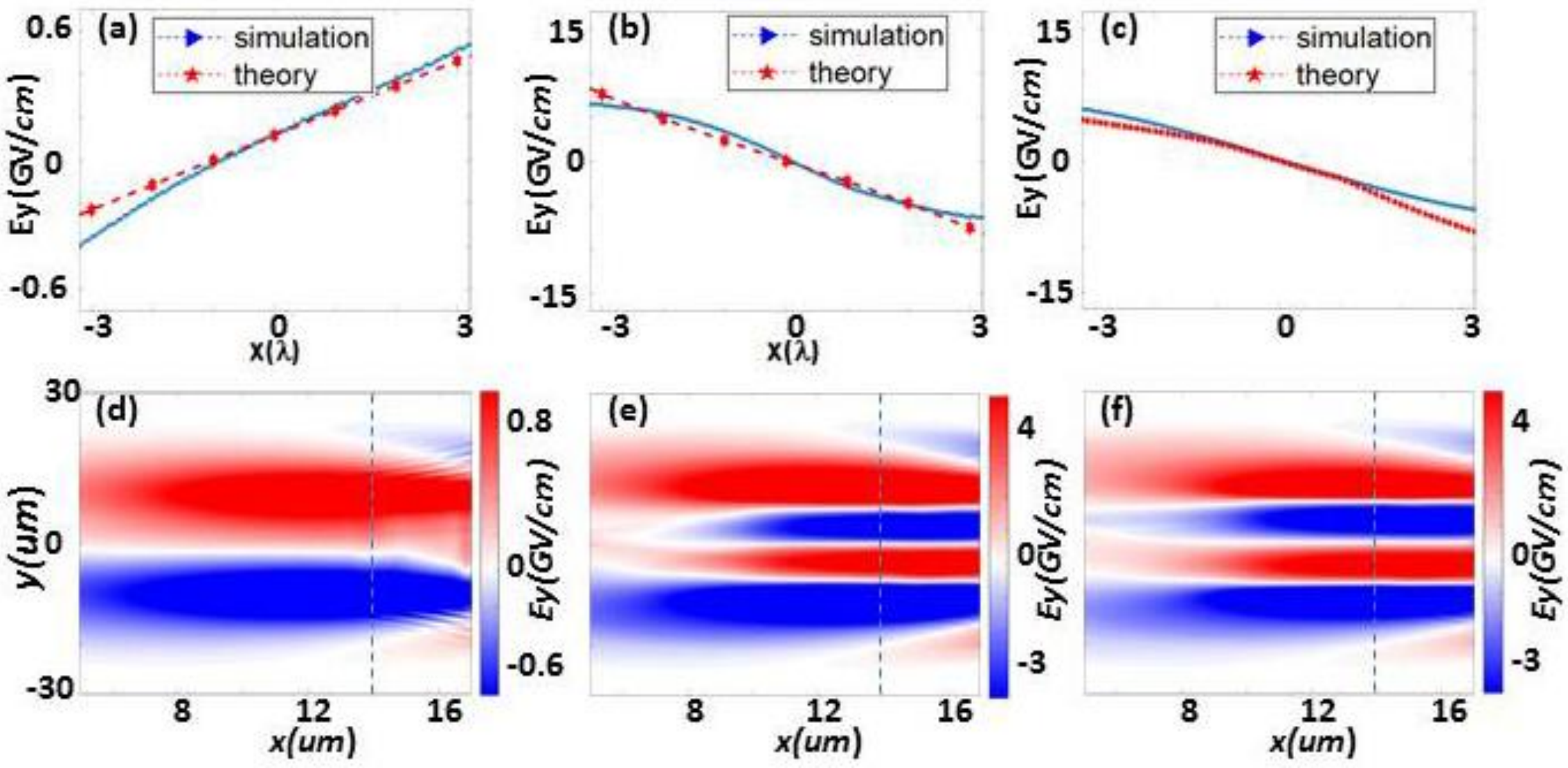

**Figure 2**

**Tight focusing proton beam with radius in nanometer scale generation based on channeled solid target**

**Q. Yu *et al.***

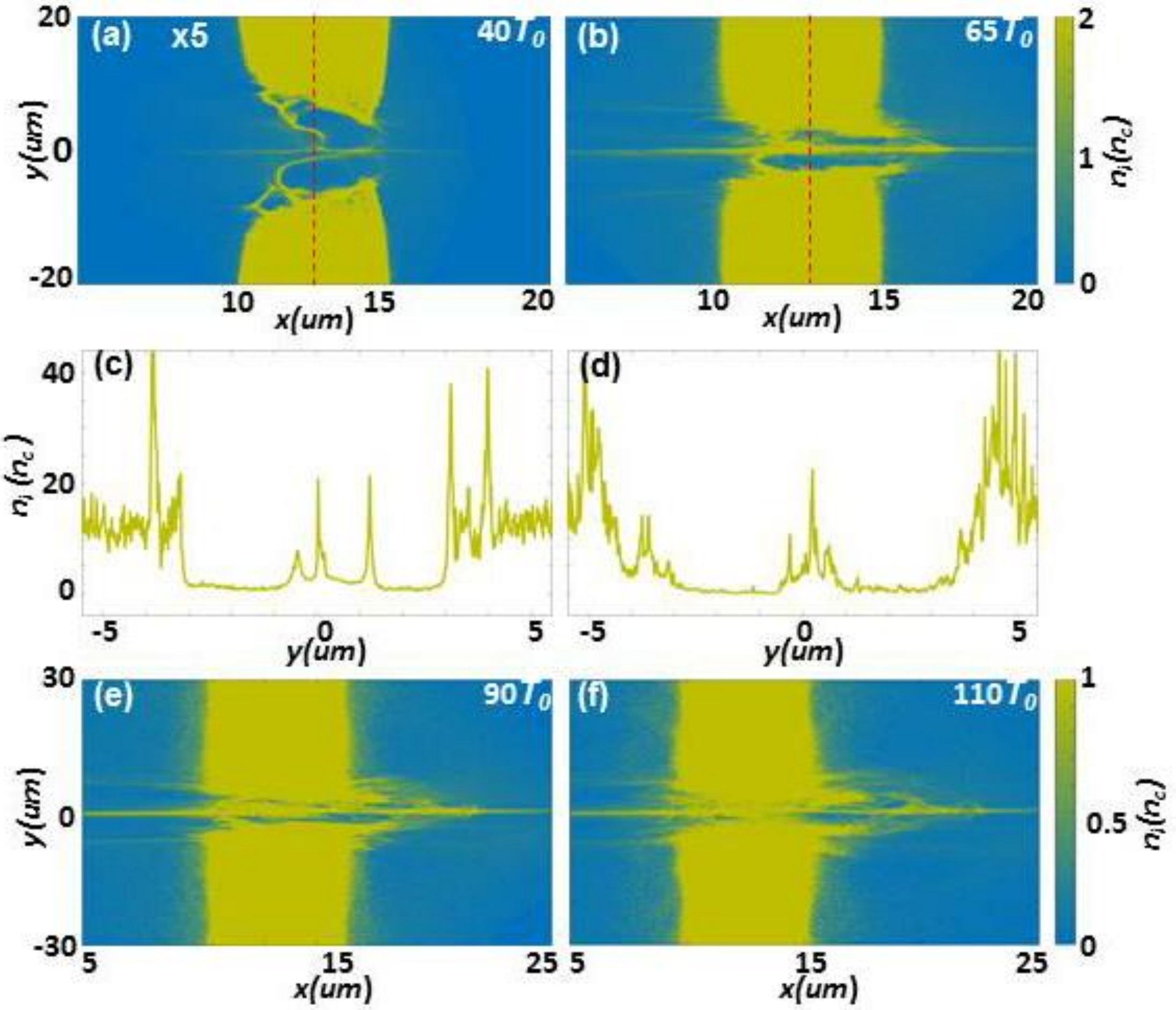

**Figure 3**

**Tight focusing proton beam with radius in nanometer scale generation based on channeled solid target**

**Q. Yu *et al.***

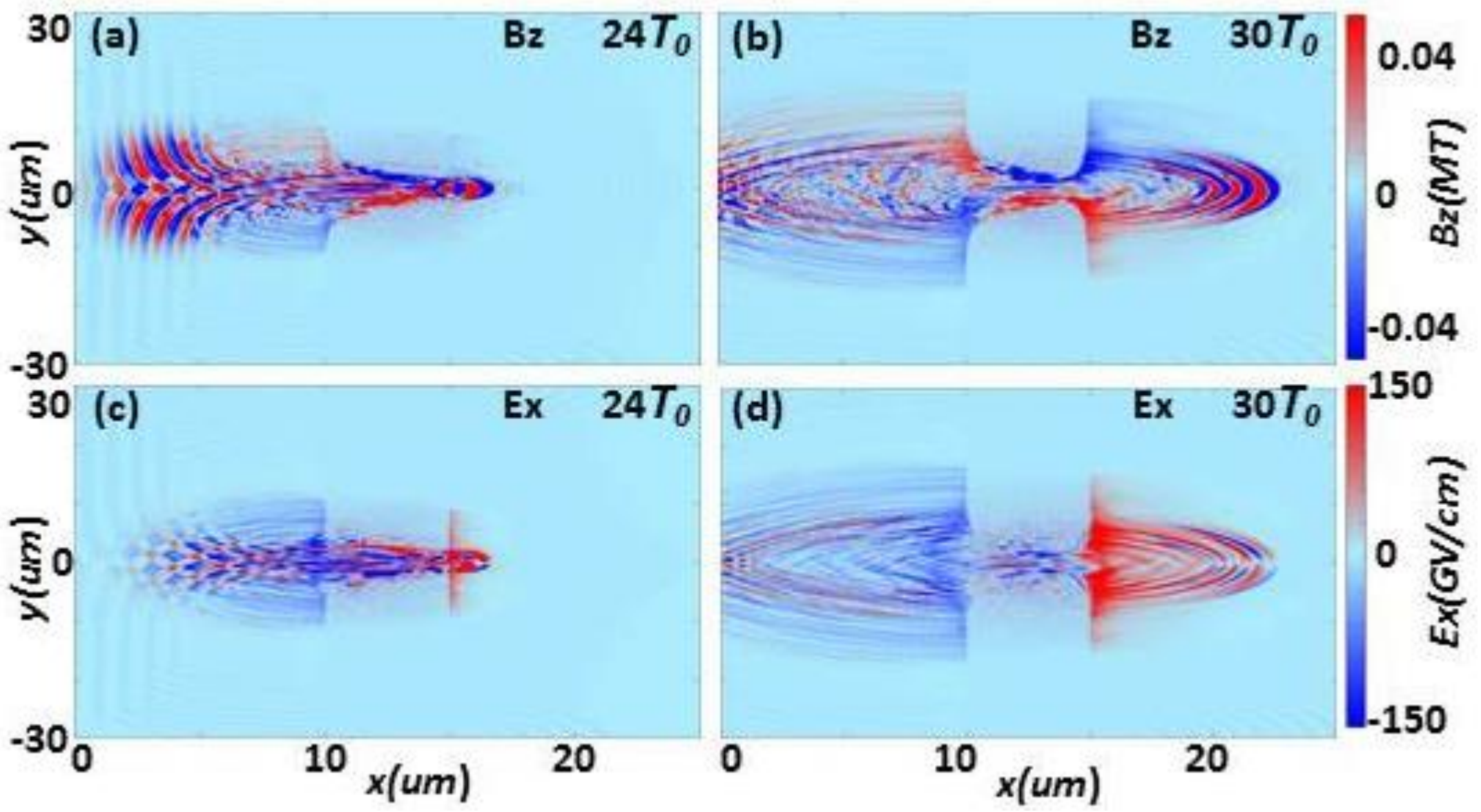

**Figure 4**

**Tight focusing proton beam with radius in nanometer scale generation based on channeled solid target**

**Q. Yu *et al.***

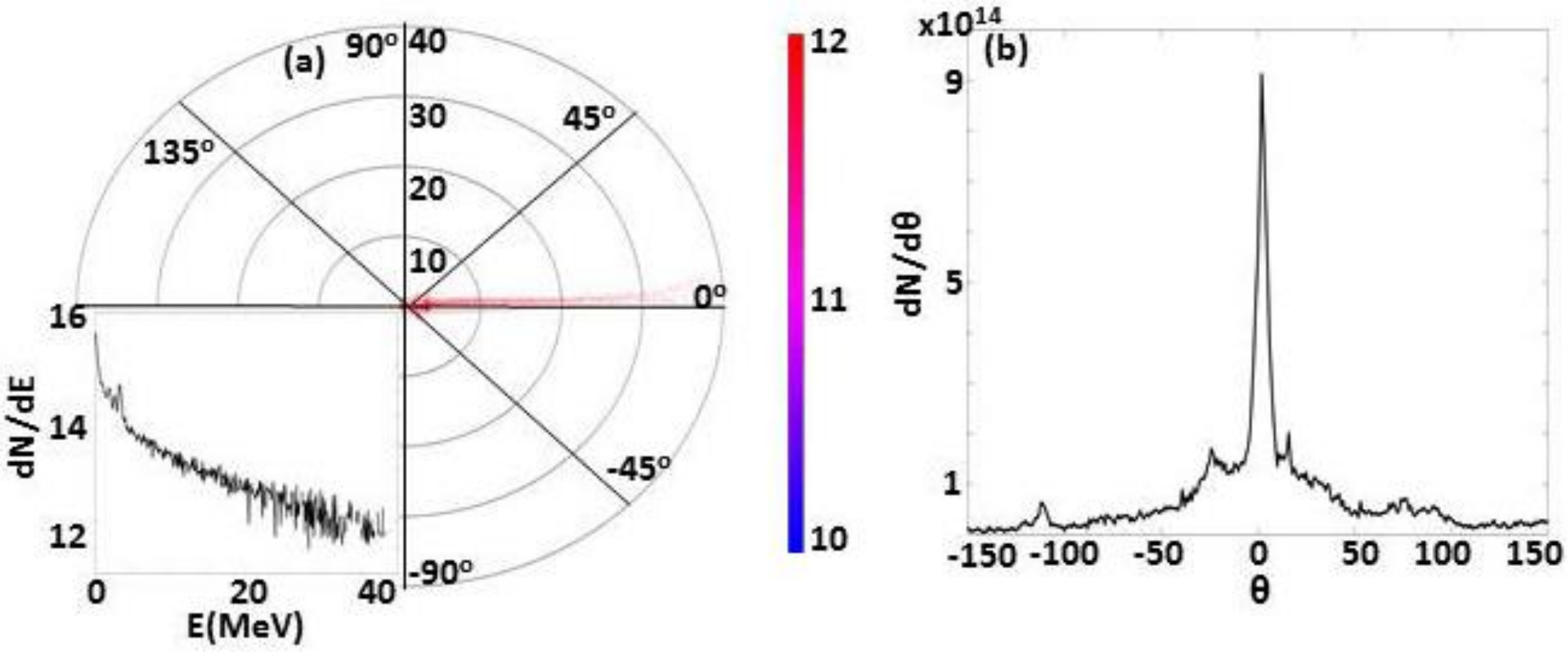

**Figure 5**

**Tight focusing proton beam with radius in nanometer scale generation based on channeled solid target**

**Q. Yu *et al.***

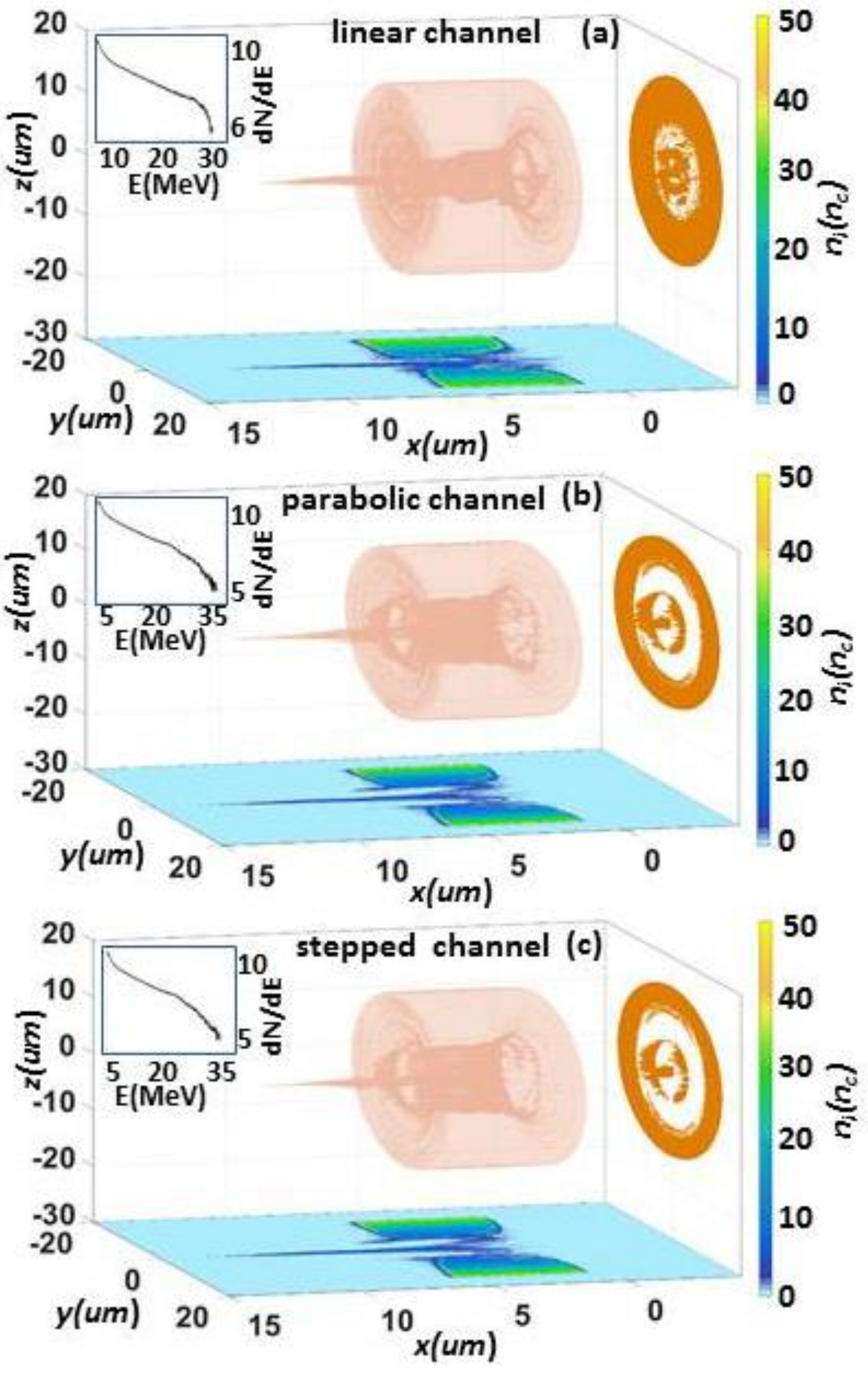